# Evidence for monopole-like topological magnetoelectric effect in image potential states

Yang Zhan[1,2], Huaiyu Zhang[1,2], Yu Wang[1,2], Chenmin Liu[1,2], Jian-Qiang Zhong[3], Yeliang Wang[4], Hong-Jun Gao[1,2], Yi-Qi Zhang[1,2], Baojie Feng[1,2], Peng Cheng[1,2*], Lan Chen[1,2*]

[1] *Institute of Physics, Chinese Academy of Sciences, Beijing, China.*

[2] *School of Physical Sciences, University of Chinese Academy of Sciences, Beijing, China*

[3] *School of Physics, Hangzhou Normal University, Hangzhou, Zhejiang, China*

[4] *MIIT Key Laboratory for Low-Dimensional Quantum Structure and Devices, School of Integrated Circuits and Electronics, Beijing Institute of Technology, Beijing, China*

*Emails: lchen@iphy.ac.cn (L. C.); pcheng@iphy.ac.cn (P. C.)

**ABSTRACT.** Magnetic monopoles, hypothetical particles behaving as isolated magnetic charges, have long been predicted by theories beyond the standard model but remain elusive in experimental detection. Subsequently, Xiaoliang Qi *et al*. proposed that magnetic monopoles can be constructed in real space by introducing an active electric field at the interface between a topological insulator and vacuum [*Science* **323**, 1184 (2009)]. Here we use scanning tunneling microscopy in the field-emission regime to realize an active electric-field geometry at the surface of the higher-order topological insulator Bi(111), and observe an anomalous splitting of the image potential states (IPSs). By tuning the dielectric properties of the substrate and film thickness, we considered that the peak-splitting of IPSs is related to the radially active electric field and topological surface states. Combined with phenomenological analysis, this peak-splitting can be attributed to the equivalent magnetic field of monopole-like topological magnetoelectric response. This work establishes field-emission IPS spectroscopy as a sensitive platform for generating active electric fields and probing the resulting image magnetic monopoles at topological surfaces.

## INTRODUCTION

The electromagnetic response of conventional insulators is linear. However, Xiao-Liang Qi and collaborators [1] have proposed that the topological insulators can host a non-linear magnetoelectric response, known as the topological magnetoelectric effect (TME). The effective action ($S_\theta$) can be written as:

$$S_\theta = \frac{\alpha\theta}{4\pi^2}\int d^3xdt\boldsymbol{E}\cdot\boldsymbol{B} \quad (1)$$

where $d^3xdt$ is the volume element of space and time, $\boldsymbol{E}$ and $\boldsymbol{B}$ are electric field and magnetic field defined within an insulator, $\alpha$ is defined as $e^2/\hbar c$, while $\theta = 0$ for conventional insulators and $\theta = \pi$ for topological insulators (TI). Previous theoretical [2,3] and experimental [4-7] studies have attempted to introduce ferromagnetic or antiferromagnetic order into topological insulators, thereby transforming the system into either quantum anomalous hall insulators or axion insulators, which manifest as integer and zero Hall conductivity plateaus in transport measurements, respectively. These works support the validity of TME effect and indirectly suggest that electric and magnetic fields may be commutative in specified condensed matter systems. More importantly, it is implied that, by theoretical symmetry, the existence of electric monopoles would permit the existence of magnetic monopoles in principle, but has not been experimentally realized till now.

Building on the TME effect, a theoretical proposal has been put forward to construct magnetic monopoles in real space [8]. The equivalent active magnetic field can be generated by applying an active electric field ($\nabla \cdot E \neq 0$) to a symmetry-broken surface (e.g. the TI-vacuum interface). Experimentally, the field emission mode of scanning tunneling microscope (STM) offers a direct method to produce such active electric field geometry, which stems from electrons are emitted from the tip and injected into the sample by applying a voltage exceeding the work function of metal tip. The attractive potential between the electron and its image charge can give rise to bound states known as image potential states (IPSs), which has been well studied by STM [9-14].

Previous studies of IPSs on the surface of three-dimensional (3D) TIs ($Bi_2Te_3$ [15] and Sb [16]) revealed that the dissipation mechanism and lifetime of IPSs are influenced by the topological surface states. However, the investigations of IPSs on the surface of higher-order topological insulators (HOTI) are absent. Bismuth is the heaviest element among stable elements, and its exceptionally strong spin-orbit coupling induces band inversion, unconventional topological states [17-21], and also has unique transport properties such as anomalous Hall [22,23] and nonlinear Hall response under zero magnetic field [24], making it an attractive platform for studying the interaction between active electric fields and topological surface states. In this letter, we measured the IPSs on Bi(111) thin film with varying thicknesses by STM, and discovered an anomalous peak-splitting phenomenon in samples beyond three bilayers. By switching the film on the substrates with different dielectric constants to alter the distribution of electric field from the tip, the peak-splitting of IPSs can be correlated with the radial component of the electric field. These results indicate that field-emission IPS spectroscopy provides a practical platform for examining how an active radial electric field couples to topological surface states. Within a phenomenological framework, the observed splitting can be discussed in terms of an effective magnetic response associated with a monopole-like TME scenario. In this sense, the present work establishes an experimental link between field-emission IPS spectroscopy and the active-field geometry envisioned in the Qi model.

## RESULTS

### IPSs on Bi(111)/Si(111) Heterojunction

We first deposited Bismuth (Bi) atoms on Si(111) to saturate the dangling bonds of Si atoms, forming a $\sqrt{3} \times \sqrt{3}$ reconstruction referred to as β-Bi [25]. Further depositing Bi atoms on this reconstructed surface results in the formation of Bi(110) islands with a black phosphorus-like structure. Once the film thickness beyond 4 bilayers (BL), the Bi(110) undergoes a structural phase transition into a Bi(111) structure [26]. In our experiments, we fabricated a high-quality Bi(111) crystalline thin film with thickness over 10 BL on Si(111) surface, as shown in Fig. 1(a). The atomic-resolution STM images of the Bi(111) surface reveal a hexagonal closed-packed arrangement with lattice constants of $a$ = 4.5Å [Fig. 1(b)]. Scanning tunneling spectroscopy (STS) measurements in Fig. 1(c) taken on the terrace exhibit a characteristic peak at 220 meV (black arrows), while at 180 meV (red arrows) near the edge, corresponding to the van Hove singularities (vHs) of surface state and edge state, respectively. Figure 1(d) and (e) recorded synchronously at 180 mV, show the topology and local density of states (LDOS) of a triangular Bi island, demonstrating that the peak is highly localized at the island's edge. The characterizations of the surface and edge states indicate that our Bi(111) film is a HOTI [18,19].

Then we increase the bias above the work function of the tip, at which point the electrons are emitted from the tip apex through field emission, inducing image charges below the sample surface. Perpendicular to the sample surface, the field-emitted electron reside in -1/*z* potential well formed by image charge, resulting in a series of bound-states, named as IPSs [16]. If the bias voltage matches the energy of an IPS, the transmission probability is abruptly enhanced and the differential conductance is risen, corresponding to multiple discrete IPS peaks in STS. First, we measured the standard IPSs spectrum on Si(111) and β-Bi, which are topologically trivial, revealing multiple independent IPS peaks shown in Fig. S1 of the Supplemental Material. The shapes of each peak conform to Lorentz distribution. According to the central limit theorem, the superposition of multiple Lorentz peaks with varied central energies converges towards Gaussian statistics. Therefore, our fitting results for each IPS peak on the Si and β-Bi surfaces accordingly conform closely to a Gaussian distribution (Fig. S1 of the Supplemental Material). The full width at half maximum (FWHM) for IPS peak exhibits a decreasing trend as energy increases, indicating that IPSs lifetimes increase with growing principal quantum number [16, 27].

In contrast to the topologically trivial samples, the IPSs spectra on the surface of Bi(111) thin films [Fig. 1(f)] show that only the $n$ = 1 IPS exhibits a standard Lorentzian line shape, while the $n$ = 2 IPS anomalously splits into two distinct subpeaks. The $n$ = 3 and 4 IPSs also display asymmetry, deviate from both Lorentzian and Gaussian distributions and possess markedly broader FWHM than the $n$ = 1 level, suggesting the IPS peaks with higher quantum number (⩾3) may also undergo split, although subpeaks are hardly distinguished due to broadening. When the energy lies above the sample work function, the electronic states are generally expected to be extended rather than form new bound-state levels. This makes it unlikely that the additional split peak arises from an independent bound state, and instead suggests that it is associated with the IPSs.

Intrinsic IPSs can also be measured experimentally through the two-photon photoemission (2PPE) technique [10, 28-30]. As no externally applied electric field is required, the 2PPE experiments provide an intrinsic representation of the IPS energy levels [12]:

$$E_n = E_{vac} - \frac{0.85eV}{(n+a)^2}, n = 1,2 \ldots\ldots \tag{2}$$

where $E_{\mathrm{vac}}$ is the vacuum energy level, and $a$ is a correction factor of the crystal field. The first three intrinsic IPSs on the surface of bismuth measured by 2PPE occur at $E$ = -3.571, -4.052, -4.170 [28]. On the other hand, in STM measurements, the metal tip introduces an extremely strong electric field of approximately $10^9$ V/m in the tunnel junction. Similar to the Shockley surface state undergoes a Stark shift under the influence of strong external electric field[31], IPSs also exhibit a comparable shift. Moreover, since the IPSs wavefunction is exposed beyond the lattice periodic potential, it is affected more intensely by the electric field [9, 16, 32]. Therefore, the IPSs in Fig. 1(f) shifts overall toward higher energy levels compared with the 2PPE results due to the Stark effect. Furthermore, the $n = 2$ IPS measured in the 2PPE experiments does no split [28]. We speculate that the peak splitting of IPSs in the STM measurements originates from the tip electric field.

During the field emission process in STM, the electric field distribution at the tunneling junction consists of two components: one originating from the metallic tip, and the other provided by the field-emitted electrons. For the tip, electrons tend to concentrate at the point of greatest curvature (the tip apex), which can be simplified as a point charge $q$. Using the mirror charge method to estimate the electric field generated by the tip, we consider an induced image charge with $q^{'} = \frac{\epsilon_r - 1}{\epsilon_r + 1} q$ ($\epsilon_r$ is the relative permittivity of the sample) within the dielectric material. The sample surface is subjected to a vertical electric field ($E_z$) and a horizontally distributed electric field ($E_r$) with central symmetry, which are respectively expressed as:

$$E_{z,tip}(z = 0) = \frac{q}{4\pi\epsilon_0} \cdot \frac{2\epsilon_r}{\epsilon_r + 1} \cdot \frac{1}{d^2} \tag{3}$$

$$E_{r,tip}(z = 0) = \frac{q}{4\pi\epsilon_0} \cdot \frac{2}{\epsilon_r + 1} \cdot \frac{r}{(d^2 + r^2)^{3/2}} \tag{4}$$

where $\epsilon_0$ is the vacuum permittivity and, $d$ is the distance between tip and sample, $r$ is the distance from the center of charge $q$. The magnitude of this electric field is regulated by the dielectric permittivity $\epsilon_r$ and the distance $d$ to the sample. Further consideration of the field emission process of electrons to be a kinetic process, a single electron emitted from the tip accelerates through the tunnelling junction (about 1 nm) under the influence of a strong electric field ($10^9$ V/m) with a travel time estimated to be about 1 femtosecond (fs). Then the electron locates on the IPSs about a few femtoseconds(*16*), and finally the current forms from the IPSs relaxing into the sample. Since the plasmon oscillation frequency $\omega_p = \sqrt{\frac{ne^2}{\epsilon_0 m^*}}$ of semiconductors, with typical carrier concentration is on the order of $10^{18}$ $cm^{-3}$, is about $10^{14}$ $s^{-1}$, we estimate the time required to establish electrostatic equilibrium is on the order of 100 femtoseconds, much greater than the electron traveling time across the junction. Therefore, the electric field of field-emitted electrons is entirely contributed by themselves. When the emitted electron is at a distance $d_e$ from the sample, the instantaneous electric field on the sample surface can be expressed as: $E_{z,e}(z = 0) =$

$\frac{e}{4\pi\epsilon_0}\cdot\frac{1}{d_e^2}$ and $E_{r,e}(z=0)=\frac{e}{4\pi\epsilon_0}\cdot\frac{r}{(d_e^2+r^2)^{3/2}}$ , which only depends on the orbital radius and lifetime of the IPSs. Both of the radial electric field $E_r$ generated by metallic tip and field-emitted electrons can be regarded as an active electric field within the *x*-*y* plane with $\nabla\cdot E_r\neq 0$.

The carrier concentration of bulk bismuth is comparable to typical semiconductors ($10^{17}$cm$^{-3}$) [33, 34], while our grown Bi(111) film with 4 nm thickness lacks sufficient free carriers to establish electrostatic equilibrium under strong external electric field. Consequently, the distribution of the electric field within the tunnel junction is primarily determined by the dielectric properties of the substrate. Considering Si(111) is a typical semiconductor substrate with a dielectric constant $\epsilon_r$ = 11.7, we give the schematic diagram of the electric field distribution in STM tunnel junction shown in Fig. 2(a). As the tunneling current increase, the tip-sample distance decreases while the tip electric field increases simultaneously, pushing all IPSs to higher energies due to Stark effects under stronger fields, as shown in Fig. 2(b). We summarize the relationship between the tunneling current and the energy of IPSs in Fig. 2(c), and the dependence of the splitting energy between $n$ = 2/2' IPS on the tunneling current in Fig. 2(d). The splitting degree of magnitude gradually increases from 0.28 eV to 0.55 eV as the tunnel junction distance decreases and the tip electric field increases. This trend is similar to the Stark shift, suggesting that the splitting also originates from the tip electric field.

### IPSs on Bi(111)/Sb(111) Heterojunction

To further understand the splitting behavior of IPSs in the electric field, we need to investigate the IPSs of Bi(111) on metal substrate. Then the high-quality Bi(111) films with thicknesses ranging from 1 to 10 BL on Sb(111) substrate are prepared. Figure 3(a) shows a large-area Bi(111) film grown on Sb(111) with the same thickness as on Si(111), and the atomic-resolution image displays a lattice constant of *a* = 4.4 Å [Fig. 3(b)], which is approximately 2% compressive strain compared with that on Si(111). Figure 3(c)-(e) present the characterization of the topological surface states and edge states of Bi(111) on Sb(111) substrate through STS and d*I*/d*V* mapping, indicating that the topological properties of the film remain unchanged under substrate-induced stress. Figure 3(f) displays the IPS peaks of Bi(111) films with different thicknesses on Sb(111). For Bi(111) films with thinner thickness (1-2 bilayers), several distinct IPS peaks corresponding to principal quantum numbers $n$ occur, similar as the situation on the topologically trivial semiconductors shown in Fig. S1 of the Supplemental Material. Upon the film thicker than three bilayers, the $n$ = 2 IPS undergoes a significant splitting and subsequently persists up to ten bilayers or beyond. It has been demonstrated that single-bilayer Bi(111) is a 2D TI [35], whereas bulk Bi(111) manifests as a 3D HOTI [18,19,21]. Theoretical DFT calculations also suggest that topological surface states on Bi(111) emerge starting from three bilayers [36]. Such thickness dependence of topological surface states also occurs in other 3D TI, such as $Bi_2Te_3$ [37] or $Bi_2Se_3$ [38]. As the film thickness decreases, the wave functions of the upper and lower surface states overlap spatially, thereby closing the energy gap or inducing a topological phase transition due to the interlayer coupling. The critical thickness at which topological surface state emerges exactly coincides with that the IPS splitting occurs, suggesting that it should correlate with the topological nature of the 3D HOTI.

As is known, being a semiconductor substrate, Si can possess an electric field with both radial and perpendicular components. In contrast, Sb as the metal substrate with $\epsilon_r \rightarrow \infty$ exhibits electrostatic shielding effects. Therefore, the radial tip electric field component $E_{r,tip} \approx 0$ and the field lines are perpendicular to the Sb surface ($E_z$) according to Eq. 3, as shown in Fig. 4(a). By adjusting tunneling current to modify the tip-sample distance $d$ and the tip electric field strength, we obtained the IPSs at different tunnelling currents ($I_t$ = 50 pA ~ 8.0 nA) on 10BL Bi(111) thin film shown in Fig. 4(b). All the IPS peaks move to higher energy as the tunnelling current increases [Fig. 4(c)], suggesting the Stark shift due to electric field component $E_z$. However, contrary to the situation in Bi(111)/Si(111) [Fig. 2(d)], the splitting energy between the $n$ = 2/2' IPS does not increase accordingly, which indicates that the splitting of IPSs is unrelated to the $E_z$. Therefore, we infer that the increasing splitting energy on Bi(111)/Si(111) is caused by the radial component of the tip electric field. Furthermore, the splitting energy of IPSs on Bi(111)/Sb(111) is approximately 0.3 eV, which is comparable to the smallest splitting energy in Bi(111)/Si(111), suggesting this value originate from transient active electric fields generated by field-emitted electrons. Furthermore, we extracted the full width at half maximum (FWHM) of $n$ = 3,4 IPSs shown in Fig.S2 of the Supplemental Material. In Fig. 2(d) and (d), we shows the trend in the variation of the FWHM is fully consistent with the splitting energy observed in $n$ = 2 IPS. This further supports our interpretation that the anomalous broadening of IPSs at high quantum numbers arises from splitting into multiple sublevels.

In addition to the radial active electric field component, there is another factor leading to the splitting of IPSs: symmetric and antisymmetric IPSs series. For a free-standing 2D material, two Rydberg series are expected to emerge due to the charges at each side of the material (symmetric state $n^+$ and antisymmetric state $n^-$). In this situation, the significant splitting sub-peaks are usually experimentally observed at IPS with $n$ = 1 [39,40]. However, the peak splitting occurs at the IPS with $n$ = 2 on Bi(111). We also measured the IPSs spectra across a step of Bi(111) island shown in Fig. S3 of the Supplemental Material, which indicates no obvious site dependence, further ruling out the possibility that this splitting is caused by confinement effects from quantum wells [41] or nanodots [14,42] formed in a film with finite thickness.

## DISCUSSION

Previous studies typically adopted a 1D quantum well model to describe the IPSs, treating them as 2D surface states confined along the $z$-axis while freely extended in the $x$-$y$ plane. In this simplification, IPSs possess only one principal quantum number $n$. However, in our experiment, we observed a new bound state of IPSs with $n$ = 2/2'. Therefore, we believe that the potential well distribution of image charge within the $x$-$y$ plane $V \propto -(d^2+r^2)^{-1/2}$ should also be taken into consideration. Similar to the central symmetric potential of the hydrogen atom model, we can define the orbital angular momentum quantum number $l$ with $n$-fold degenerate. When an electron is in a centrally symmetric potential field, its energy levels corresponding to different quantum numbers $l$ degenerate and do not manifest as splitting peaks of IPS. However, if the electron is under a magnetic field, the intrinsic magnetic moment $\mu_B$ carried by $l$ gains additional energy and undergoes a splitting, resulting in splitting energy given as $\Delta E_{split} = \mu_B B_z \sqrt{l(l+1)}$, $l$ = 0 ... $n$ - 1. We performed line fitting on the IPSs with different $n$ using quantum numbers $l$ and the results are shown in Fig. 5(a). All the IPSs peaks are well refitted, especially the $n$ = 3 and 4 IPSs with asymmetry shape, suggesting the presence of a built-in magnetic field.

Because the onset of IPS splitting closely tracks the thickness regime in which Bi(111) develops topological surface states, it is natural to discuss the observations within a TME response scenario. According to the tip electric field distribution in a vacuum-TI-semiconductor heterojunction [Fig. 5(c)], the electric field distribution on the TI surface can be regarded as an active electric field $E_r$ with center symmetry in the $x$-$y$ plane, which is shown in Fig. 5(d) and (e). Under the influence of the TME effect, an active electric field generates Hall currents $j_\theta$ centered around the charge. Each Hall current loop at radius $r$ generates an equivalent magnetic field perpendicular to the surface at the center with magnitude $B_z(r) = \mu_0 \frac{\sigma_{xy} E(r)}{2r}$. Integrating all Hall currents within the *x*-*y* plane yields the total equivalent magnetic field $B_z$:

$$B_z \propto \frac{1}{\epsilon_r + 1} \frac{1}{d^2}$$

An intrinsic orbital magnetic moment $\mu_B$ of IPS acquires additional energy under the influence of a magnetic field, resulting in an energy level difference $\Delta E_{split} \propto \mu_B B_z \propto d^{-2}$. By further extrapolating the model theoretically, we can derive the attenuation of the equivalent magnetic field generated by the Hall circulation in the direction perpendicular to the sample (*z*-axis): $B(z) = (z + d)^{-2}$. If $d << z$, an equivalent magnetic field is proportional to $z^{-2}$ at the center of the Hall circulation, identical to the active magnetic field from a monopole-like TME effect. We can further estimate that the splitting magnitude of the IPSs energy levels induced by such an image magnetic monopoles in hydrogen-like atoms is approximately 200 meV (see Supplementary Materials 4), comparable to the experimental splitting value of IPSs on Bi(111)/Sb(111).

In field emission mode, the relationship between the tunneling current *J* and electric field $E_z$ can be described by the Fowler-Normheid equation (FN equation) $J = \alpha E_z^2 e^{-\beta / E_z}$, where the fitting parameter $\alpha,\ \beta$ is jointly determined by factors such as the tip shape and the work function of tip and sample. The emitted electrons and their mirror charges form an electric dipole pair, whose intrinsic dipole moment *p* undergoes a Stark shift under the electric field $E_z$. Consider a pair of opposite charges confined at a distance of 1 nm and subjected to a strong electric field of $10^9$ V/m, their electric dipole moment shifts by approximately 1 eV under the Stark effect, consistent with the observed magnitude in our experiments.

The interactions between IPSs and the electric field are approximately dominated by the first-order Stark effect, inducing an energy shift of $\Delta E_{stark} \propto p_n E_z \propto d^{-2}$ (from Eq. 3). Therefore, both Stark shift of each IPS and the splitting energy between the *n* = 2/2' IPS sublevels on Bi(111)/Si(111) are expected to follow the same variation trends as the tip-sample distance. The relationship between the Stark shift as well as splitting energy between *n* = 2/2' IPS with the tunneling current should follow the FN equation, and the fitting result are shown in Fig. 2(c) and (d). Then, we extracted the Stark shift and splitting energy between *n* = 2/2' IPS at different tip-sample distances on the Bi(111)/Si(111) surface and plotted them in Fig. 5(b), confirming a linear relationship. This correlation indicates that $\Delta E_{split} \propto \Delta E_{stark} \propto d^{-2}$, consistent with the mechanism described above and further supported our speculations.

Since topological insulators thin films contain two interfaces with the vacuum and the substrate, the $z$-direction electric field penetrating the topological insulator generates Hall currents at both interfaces. As the film thickness approaches few monolayers, the states of the two surfaces overlap spatially, inducing the Hall currents on the two interfaces to cancel each other out. This process is consistent with the physical mechanism whereby reducing the 3D TI thickness induces hybridization of the top and bottom surface states, thereby opening a topologically trivial band gap. Consequently, for Bi(111) thin films with thickness less than three monolayer, IPSs do not exhibit splitting [Fig. 3(f)]. Within a phenomenological framework, these observations can be understood in terms of an effective magnetic response associated with a monopole-like TME scenario.

## CONCLUSION

In summary, we observed an anomalous splitting of image potential states on Bi(111) thin films. By tuning the dielectric properties of the substrate and the film thickness, we found that the splitting is strongly correlated with the radial component of the active electric field and with the emergence of topological surface state. Based on phenomenological models, we propose that the radial electric field may generate an equivalent magnetic field through TME effect, which in turn contributes to the splitting of IPSs sublevels. More importantly, our results show that IPS spectroscopy naturally provides an experimental route for generating the active radial electric-field geometry envisioned in the Qi model and for probing the resulting surface response spectroscopically. Although there are currently no effective experimental methods to detect whether field-emitted electrons actually induce magnetic fields near topological insulator surface, the anomalous splitting of IPS states may provide an indirect evidence for the existence of a monopole-like topological magnetoelectric effect.

## METHODS

### Sample preparation

The experiments were performed in a home-built low temperature STM system equipped with a molecular beam epitaxial (MBE) chamber, with base pressure of $3\times10^{-10}$ mbar. High-purity Bi (99.999%) was evaporated from a homemade evaporation source. The silicon wafer is n-type doped with a resistivity of $0.05\ \Omega\cdot cm$. Si(111) wafer was flashed to about 1200 ℃ for several times to obtain a 7×7 reconstruction as the substrate. Clean Sb(111) surface was achieved through Ar ion sputtering and then annealing at 400°C for several cycles. The √3-Bi reconstruction was prepared by depositing a monolayer of Bi atoms onto Si(111) at room temperature, followed by annealing at 380℃. Bi(111) film on both Si(111) and (111) was prepared by room temperature deposition, with the deposition rate is about 0.3 ML/min.

**STM measurements**

The STM characterizations were performed at liquid-helium temperature with a tungsten tip. STM images were obtained in the constant-current mode. The *dI/dV* signals were recorded by standard lock-in detection with a modulation of 20 mV at a frequency of 931 Hz while sweeping the sample bias in an open feedback loop configuration. When probing IPS, the STM operates in constant-current mode, where the feedback loop controls the tip-sample distance $z$ to maintain a constant current $I_t$. As the bias voltage is increased, the tip is gradually pulled away from the sample, which can compensate for the increasing of Stark shift due to the higher voltage.

## Acknowledgements

This work was financially supported by the Ministry of Science and Technology (MOST) of China (Grant Nos. 2024YFA1409100), the National Natural Science Foundation of China (Grant Nos.T2325028, 12134019), and L.C. acknowledges the support from the CAS Project for Young Scientists in Basic Research (Grant No. YSBR-054).

## Data availability

The main data supporting the finding of this work are available within this paper. Extra data are available upon reasonable request from the corresponding authors.

## Figures and captions

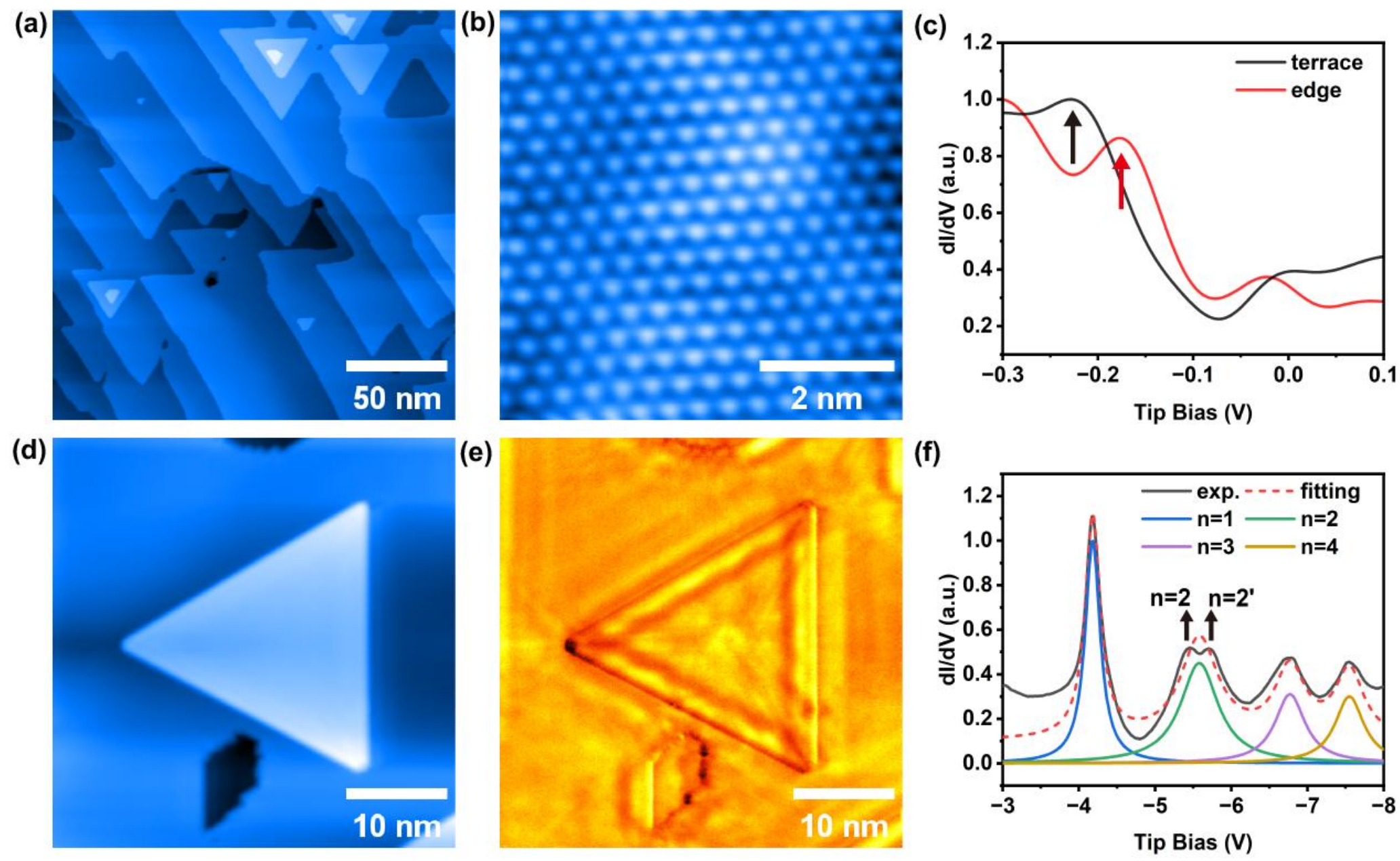

Figure 1. IPSs on HOTI Bi(111)/Si(111) thin film. (a) Topographic image of 10-BL Bi(111) thin film on Si(111) substrate (200×200 $nm^2$). (b) Atomical-resolved STM topographic image (6×6 $nm^2$). (c) STS on the terrace away from the edges (black line) and near the edge (red line) (setpoint: $V_t$ = -0.3 V, $I_t$ = 100 pA). Red arrows and black arrows indicate the vHS of surface states and edge states, respectively. (d) Topographic image of one triangular Bi(111) island. (e) d$I$/d$V$ mapping simultaneously acquired with the topographic image (d) (setpoint: $V_t$ = -180 mV, $I_t$ = 100 pA), showing high conductance at each edge of the triangle island. (f) IPSs measured on the terrace away from the edges (setpoint: $V_t$ = -2V, $I_t$ = 200 pA). Fitting the four IPSs using Lorentzian distributions reveals that only the $n$ = 1 IPS peak exhibits good agreement with Lorentzian statistics. The $n$ = 2 IPS splits into two sublevels 2/2', while the $n$ = 3/4 IPSs displays asymmetric characteristics, significantly deviating from Lorentzian statistics.

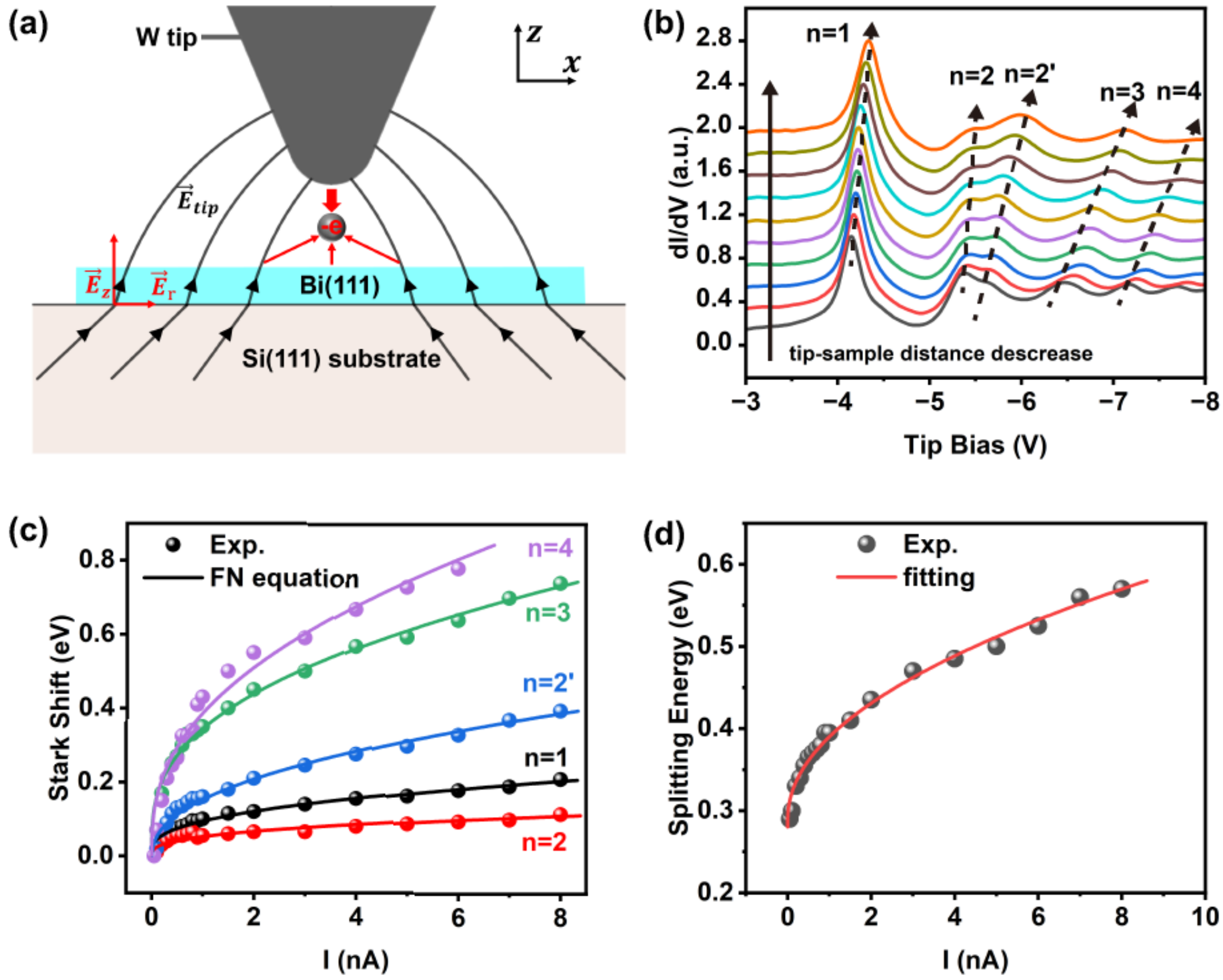

Figure 2. Electric field dependence of IPSs on Bi(111)/Si(111). (a) Schematic diagram of the electric field distribution in a metal-vacuum-semiconductor tunnel junction. Metallic tip and semiconductor substrate achieve electrostatic equilibrium, contributing to a radial active electric field ($E_r \neq 0$). The emitted electrons also contribute an instantaneous active electric field. (b) IPSs recorded at different set points ($V_t$ = -2 V, $I_t$ = 50 pA ~ 8.0 nA). Each spectral line is vertically shifted by 0.1 (a. u.). (c) Energy shifts of IPSs with the different tunneling current due to the Stark effect. (d) The splitting energy between $n$ = 2/2' IPS subpeaks varies with the tunnelling current. The splitting energy continuously increases from 0.28 eV to 0.55 eV. The points in (c) and (d) were all fitted using the FN equation.

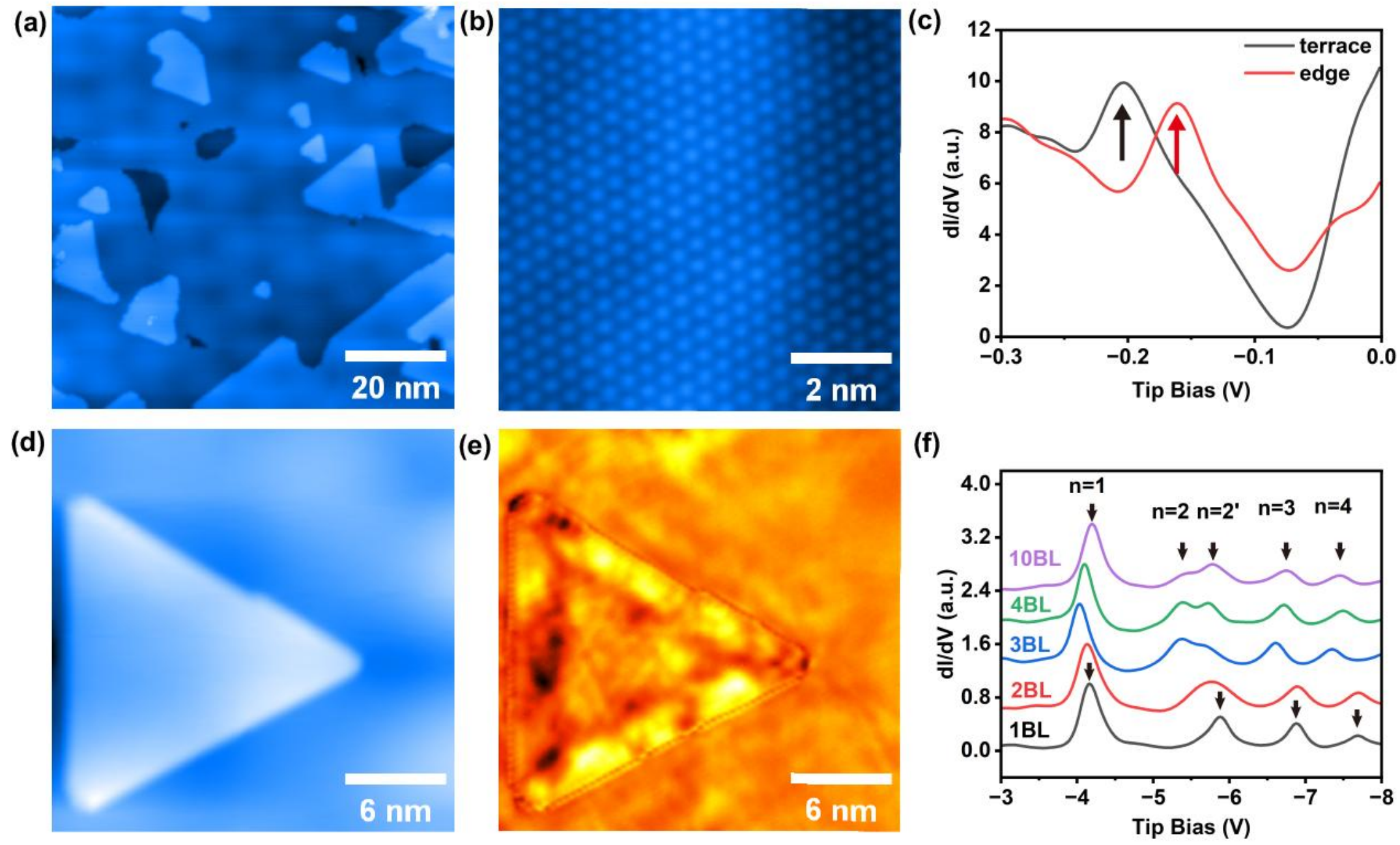

Figure 3. IPSs on HOTI Bi(111)/Sb(111) thin film. (a) Topographic image of 10-BL Bi(111) thin film on Sb(111) substrate (80×80 $nm^2$). (b) Atomical-resolved STM topographic image (8×8 $nm^2$). (c) STS on the terrace away from the edges (black line) and near the edge (red line) (setpoint: $V_t$ = -0.3 V, $I_t$ = 100 pA). Red arrows and black arrows indicate the vHS of the surface states and edge states, respectively. (d) Topographic image of one triangular Bi(111) island. (e) d$I$/d$V$ map simultaneously acquired with the topographic image (d) (setpoint: $V_t$ = -170 mV, $I_t$ = 100 pA), showing high conductance at each edge of the triangle island. (f) IPSs signals measured on Bi(111) films of different thicknesses (setpoint: $V_t$ = -2 V, $I_t$ = 8.0nA), each spectral line is vertically shifted by 0.6 (a. u.). When the film thickness exceeds 3 bilayers, the IPS with $n$ = 2 begins to split into two sublevels.

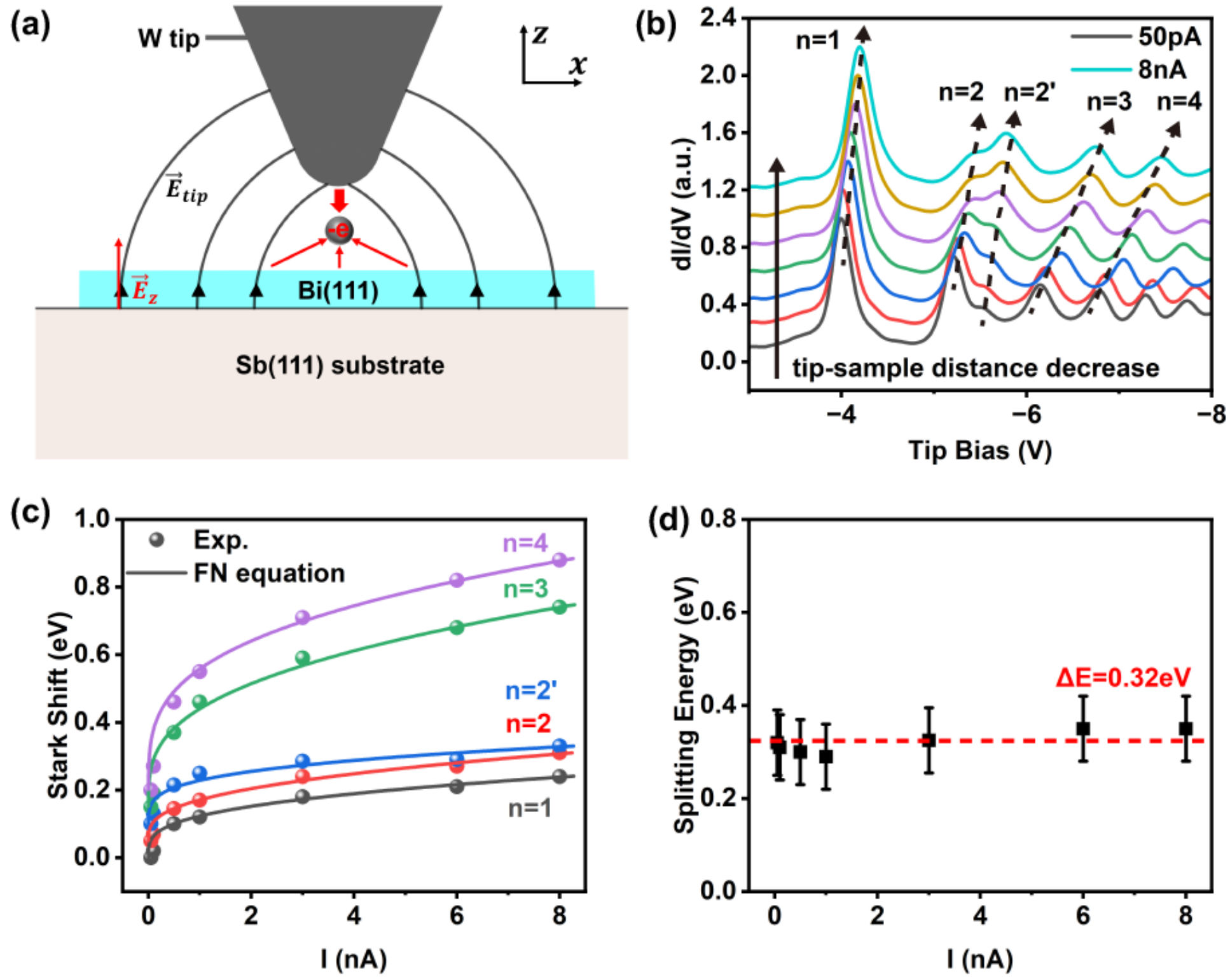

Figure 4. Electric field dependence of IPSs on Bi(111)/Sb(111). (a) Schematic diagram of the electric field distribution in a metal-vacuum-metal tunnel junction. Compared with electric field distribution on semiconductor substrates, the more metallic Sb substrate shields the radial active electric field $E_r$. Since field emission of electrons is an ultrafast process that cannot establish electrostatic equilibrium with the substrate, the active electric field of the electrons cannot be effectively blocked. (b) IPSs recorded at different set points ($V_t$ = -2 V, $I_t$ = 50 pA ~ 8.0 nA), each spectral line is vertically shifted by 0.1 (a. u.). (c) Energy shift of IPSs at different tunneling currents due to the Stark effect, the points were fitted using the FN equation. (d) The splitting energy $\Delta E = E_{n=2} - E_{n=2'}$ varies with the different tunnelling currents, revealing the splitting energy difference remains nearly constant.

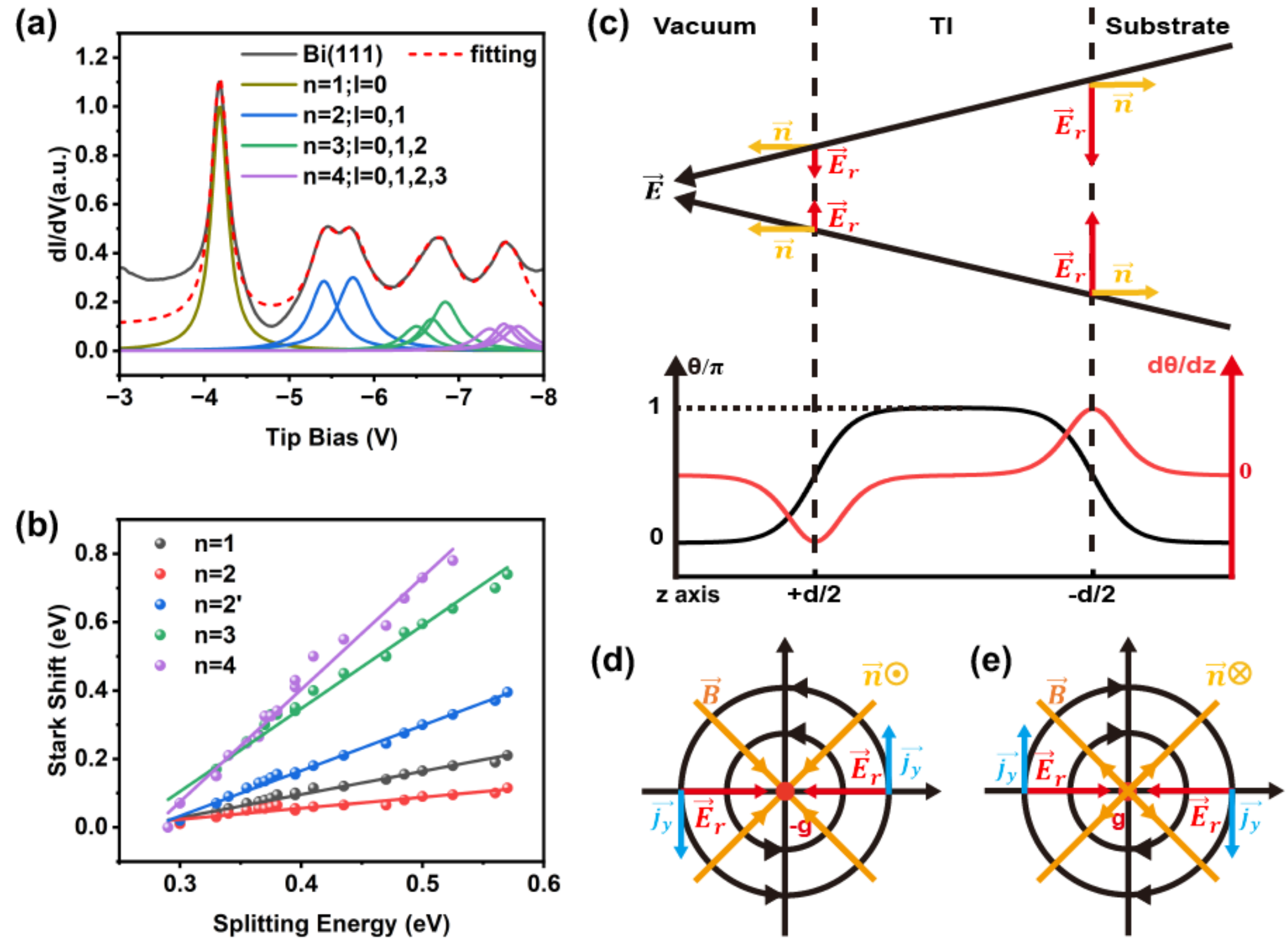

Figure 5. Theoretical Model of TME Effect Induced Equivalent Magnetic Field. (a) IPSs on the 10BL Bi(111)/Si(111) surface, refitting of IPS spectra with orbital angular momentum quantum number *l*. (b) Linear relationship of the Stark shift of each IPS and the splitting energy between the 2/2' sublevels. (c) Vacuum-TI-substrate sandwich structure. Vacuum and substrate represent topologically trivial ($\theta = 0$), the central TI represent topologically nontrivial ($\theta = \pi$). An applied electric field at the interface between topological insulators and topologically trivial materials generates a Hall current, which can be expressed as $J_{Hall} = (\partial_z \theta) \times E(r)$. (d), (e) Top view of Hall circulation at two surfaces of the TI ($z = \pm d/2$), and equivalent active magnetic field is generated at the center of the circulation.